\documentclass[11pt,a4paper]{article}

\usepackage{amsmath,amsfonts,amssymb,bm}
\usepackage{graphicx}
\usepackage{url}
\usepackage{float}
\usepackage{enumerate}

\usepackage[text={14cm,21cm},centering]{geometry}
\renewcommand{\baselinestretch}{1.2}

\usepackage{amsthm}

\theoremstyle{plain}

\newtheorem{alg}{Algorithm}[section]

\theoremstyle{definition}
\newtheorem{notation}{Notation}

\def\EE{\mathbb{E}}

\def\Em{\mathbb{E}}
\def\Var{\mathbb{V}ar}

\newcommand{\simindist}{\stackrel{\rm d}{\sim}}

\def\Zset{\mathbb{Z}}

\title{Counting with Combined Splitting and Capture-Recapture Methods}
\author{%
Paul Dupuis$^{a}$\\
{\it Brown University, Providence, USA}\\
\and 
Bahar Kaynar$^{b}$ $\quad$ Ad Ridder$^{d}$\\
{\it Vrije University, Amsterdam, Netherlands}\\
\and 
Reuven Rubinstein$^{c}$ $\quad$ Radislav Vaisman\\
{\it Technion, Haifa, Israel}
}

\begin{document}
\maketitle 
\footnotetext[1]{Research supported by AFOSR grant FA9550-09-0378.}
\footnotetext[2]{Research supported by NWO grant 400-06-044.} 
\footnotetext[3]{Research supported by BSF(Binational
Science Foundation) grant 2008482, and by NWO grant 040-11-168.}
\footnotetext[4]{Corresponding author. Department of Econometrics and Operations Research; 
Vrije University Amsterdam; Netherlands.  Email address:
\url{aridder@feweb.vu.nl}}

%
\begin{abstract}
We apply the splitting method to three well-known counting
problems, namely 3-SAT, random graphs with
prescribed degrees, and binary contingency tables.
We present an enhanced version of the splitting method
based on the capture-recapture technique,
and show by experiments the superiority of this technique
for SAT problems in terms of variance of the associated estimators,
and speed of the algorithms.
\end{abstract}


\section{Introduction}\label{s:intro}
In this paper we apply the splitting method introduced in
\cite{BotevKroese11} to a variety of counting problems in \#P-complete.
Formally, given any decision problem in the class NP, e.g.\ the satisfiability
problem (SAT), one can
formulate the corresponding counting problem which asks for the
total number of solutions for a given instance of the problem.
In the case of the SAT problem, this corresponding counting problem
has complexity \#SAT. Generally, the complexity class \#P consists
of the counting problems associated with the decision problems
in NP. Clearly, a \#P problem is at least as hard as its corresponding
NP problem. In this paper we consider \#P-complete problems.
Completeness is defined similarly as for the decision problems:
a problem is \#P-complete if it is in \#P, and if every \#P problem
can be reduced to it in polynomial counting reduction.
This means that exact solutions to these problems cannot be obtained
in polynomial time, and accordingly, our
study focuses on approximation algorithms.
For more background on the complexity theory of problems we refer
to \cite{Papadimitriou94}.
\par
The proposed splitting algorithm for approximate counting is a
randomized one. It is based on designing a sequential
sampling plan, with a view to decomposing a ``difficult'' counting
problem defined on some set $\mathcal{X}^*$ into a number of
``easy'' ones associated with a sequence of related sets
$\mathcal{X}_0, \mathcal{X}_1,\ldots,\mathcal{X}_m$ and such that
$\mathcal{X}_m = \mathcal{X}^*$. Splitting algorithms explore the
connection between counting and sampling problems, in particular
the reduction from approximate counting of a discrete  set to
approximate sampling of elements of this set, with the sampling
performed, typically, by some Markov chain Monte Carlo method.
\par
Recently, counting problems have attracted research interest,
notably the so-called model counting or \#SAT, i.e. computing the
number of models for a given propositional formula \cite{Gomes08}.
Although it has been shown that many solution techniques for SAT
problems can be adapted for these problems, yet due to the
exponential increase in memory usage and running times of these
methods, their application area in counting is limited.
This drawback motivated the approximative approach mentioned earlier.
There are two main heuristic
algorithms for approximate counting methods in \#SAT. The first one,
called \verb"ApproxCount", is introduced by Wei and Selman in
\cite{WeiSelman05}. It is a local search method that
uses Markov Chain Monte Carlo (MCMC) sampling to compute an
approximation of the true model count of a given formula. It is fast and has been
shown to provide good estimates for feasible solution counts, but,
in contrast with our proposed splitting method,
there are no guarantees as to the uniformity of the MCMC samples.
Gogate and Dechter \cite{GogateDechter07} recently proposed a
second model counting technique called \verb"SampleMinisat", which
is based on sampling from the so-called backtrack-free search space
of a Boolean formula through \verb"SampleSearch".
An approximation of the search tree thus found is used
as the importance sampling density instead
of the uniform distribution over all solutions. Experiments with
\verb"SampleMinisat" show that  it is very fast and typically it
provides  very good estimates.
\par
The splitting method discussed in this work  for counting in
deterministic problems is based on its classic counterpart
for efficient estimation of rare-event probabilities in
stochastic problems. The relation between rare-event simulation
methods and approximate counting methods have also been discussed,
for instance, by
Blanchet and Rudoy \cite{BlanchetRudoy09},
Botev and Kroese \cite{BotevKroese08}, and Rubinstein \cite{Rubinstein10};
see also \cite[Chapter 9]{RubinsteinKroese08}.
\par
As said, we propose to apply the sequential sampling method
presented in \cite{BotevKroese11} which  yields a
product estimator for counting the number of solutions
$|\mathcal{X}^*|$, where the product is taken over the estimators of
the consecutive conditional probabilities, each of which
represents  an ``easy'' problem. In addition, we shall consider
an alternative version, in which we use the generated
samples after the last iteration of the splitting algoritm
as a sample for the capture-recapure method. This method
gives us an alternative estimate of the counting problem.
Furthermore, we shall study an extended version of the
capture-recapture method when the problem size is too
large for the splitting method to give reliable
estimates. The idea is to decrease artificially the problem size
and then apply a backwards estimation.
Whenever applicable, the estimators associated with our proposed
enhancements outperform the splitting estimators in terms
of variance.
\par
The paper is organized as follows.
We first start with describing the splitting method in detail
in Section \ref{s:split}.
Section \ref{s:CapRecap} deals with the combination of the
classic capture-recapture method with the splitting algorithm.
Finally, numerical results and concluding remarks are presented
in Sections \ref{s:numerical} and \ref{s:conc}, respectively.

\section{Splitting Algorithms for Counting}\label{s:split}
The splitting method is one of the main techniques for the efficient
estimation of rare-event probabilities in stochastic problems. The
method is based on the idea of restarting the simulation in certain
states of the system in order to obtain more occurrences of the rare
event. Although the method originated as a rare event simulation
technique (see \cite{AsmussenGlynn07}, \cite{Ecuyer07}, \cite{Garvels00},
\cite{GarvelsRubinstein10},
\cite{Lagnoux09}, \cite{Melas97}), it has been modified
in \cite{BlanchetRudoy09}, \cite{BotevKroese08}, and \cite{Rubinstein10},
for counting and combinatorial optimization problems.
\par
Consider a NP decision problem with solution set $\mathcal{X}^*$,
i.e., the set containing all solutions to the problem.
We are interested to compute the size $|\mathcal{X}^*|$
of the solution set.
Suppose that there is a larger set $\mathcal{X}\supset\mathcal{X}^*$
which can be represented by a simple description or formula;
specifically, its size $|\mathcal{X}|$ is known and easy to compute.
We call $\mathcal{X}$ the state space of the problem.
Denote by $p=|\mathcal{X}^*|\,/\,|\mathcal{X}|$
the fraction (or ``probability'') of the solution set
w.r.t.\ the state space. Since $|\mathcal{X}|$ is known, it suffices
to compute $p$. In most cases $p$ is extremely small, in other words
we deal with a rare-event probability. However, assuming we can
estimate $p$ by $\hat{p}$, we obtain automatically
\[
\widehat{|\mathcal{X}^*|}= |\mathcal{X}| \hat{p}
\]
as an estimator of $|\mathcal{X}^*|$. Note
that straightforward simulation based on generation of
i.i.d.\  uniform samples $X_i\in\mathcal{X}$ and delivering
the Monte Carlo estimator
$\hat{p}_{\rm MC}=\frac1N \sum_{i=1}^N I_{\{X_i\in\mathcal{X}^*\}}$ as an
unbiased estimator of $|\mathcal{X}^*|/|\mathcal{X}|$ fails
when $p$ is a rare-event probability. To be more specific, assume a
parametrization of the decision problem. The size of
the state space $|\mathcal{X}|$ is parameterized by $n$,
such that $|\mathcal{X}|\to\infty$ as $n\to\infty$.
For instance, in SAT $n$ represents the number of variables.
Furthermore we assume that the fraction of the solution set $p\to 0$
as $n\to\infty$. The required sample size $N$ to obtain a
relative accuracy $\varepsilon$ of the 95\% confidence interval by
the Monte Carlo estimation method is \cite[Chapter 6]{AsmussenGlynn07}
\[
N\approx \frac{1.96^2}{\varepsilon^2 p},
\]
which increases like $p^{-1}$ as $n\to\infty$.
\par
The purpose of the splitting method is to estimate $p$ more
efficiently via the following steps:
\begin{enumerate}[1.]
\item
Find a sequence of sets $\mathcal{X} =\mathcal{X}_0,
\mathcal{X}_1,\ldots,\mathcal{X}_m$ such that $\mathcal{X}_0 \supset
\mathcal{X}_1 \supset \cdots \supset \mathcal{X}_m = \mathcal{X}^*$.
\item
Write $|\mathcal{X}^*| = |\mathcal{X}_m|$ as the telescoping product
\begin{equation}\label{e:telescoping}
|\mathcal{X}^*| = |\mathcal{X}_0| \prod_{t=1}^m \frac{|
\mathcal{X}_t|}{|\mathcal{X}_{t-1}|},
\end{equation}
thus the target probability becomes a product $p = \prod_{t=1}^m
c_t$, with ratio factors
\begin{equation}\label{e:ratiofactor}
c_t = \frac{|\mathcal{X}_t|}{|\mathcal{X}_{t-1}|}.
\end{equation}
\item
Develop an efficient estimator $\hat{c}_t$ for each $c_t$ and
estimate  $|\mathcal{X}^*|$ by
\begin{equation}\label{e:prodestimator}
\hat{\ell} = \widehat{|\mathcal{X}^*|} = |\mathcal{X}_0|\, \hat{p}=
|\mathcal{X}_0|\, \prod_{t=1}^m \hat{c}_t.
\end{equation}
\end{enumerate}
\noindent
It is readily seen that in order to obtain a meaningful
estimator of $|\mathcal{X^*}|$, we have to solve the following two
major problems:
\begin{enumerate}[(i).]
\item
Put the counting problem into the framework \eqref{e:telescoping} by
making sure that
\begin{equation}\label{e:decreasingsets}
\mathcal{X}_0 \supset \mathcal{X}_1 \supset \cdots \supset \mathcal{X}_m =
\mathcal{X}^*,
\end{equation}
such that each $c_t$ is not a rare-event probability.
\item
Obtain a low-variance estimator $\hat{c}_t$ of each ratio $c_t$.
\end{enumerate}
\noindent
To this end, we propose an adaptive
version of the splitting method. As a demonstration,
consider a specific family of decision problems,
namely those whose solution set is finite and given by linear
integer constraints.
In other words, $\mathcal{X}^*\subset\Zset_+^n$
is given by
\begin{equation}\label{e:ipset}
\begin{cases}
\sum_{j=1}^n a_{ij}x_j = b_i,  & \quad i=1, \ldots, m_1;\\
\sum_{j=1}^n a_{ij}x_j \ge b_i, & \quad i=m_1+1, \ldots, m_1+m_2=m;\\
x_j \in \{0,1,\ldots,d\},  & \quad \forall j=1, \ldots, n.
\end{cases}
\end{equation}
Our goal is to  count the number of feasible solutions (or points)
to the set \eqref{e:ipset}. Note that we  assume  that we know, or can
compute easily, the bounding finite set
$\mathcal{X}=\{0,1\ldots,d\}^n$, with points
$\bm{x}=(x_1,\ldots,x_n)$ (in this  case
$|\mathcal{X}| = (d+1)^n$) as well
for other counting problems.
\par
Below we follow \cite{Rubinstein10}.
Define the Boolean functions $C_i: \mathcal{X}\to\{0,1\}$
($i=1,\ldots,m$) by
\begin{equation}\label{e:booleanfuncs}
C_i(\bm{x}) = \begin{cases}
I_{ \{\sum_{j=1}^n a_{ij} x_j=b_i\}}, & \quad i=1,\ldots,m_1;\\
I_{\{ \sum_{j=1}^n a_{ij} x_j\ge b_i\}} , & \quad i=m_1+1,\ldots,
m_1+m_2.
\end{cases}
\end{equation}
Furthermore, define the function $S: \mathcal{X}\to\Zset_+$ by
counting how many constraints are satisfied by a point
$\bm{x}\in\mathcal{X}$, i.e.,
$S(\bm{x})= \sum_{i=1}^m C_i(\bm{x})$.
Now we can formulate the counting problem as a probabilistic problem
of evaluating
\begin{equation}\label{e:problemprob}
p = \Em_{f} \left  [ I_{ \{S(\bm{X})= m \} } \right] ,
\end{equation}
where $\bm{X}$ is a random point on $\mathcal{X}$, uniformly
distributed with probability density function (pdf)
$f(\bm{x})$, denoted by $\bm{X}\simindist f=\mathcal{U}(\mathcal{X})$.
Consider an increasing sequence of thresholds
$0=m_0<m_1<\cdots<m_{T-1}<m_T=m$, and define
the sequence of  decreasing sets \eqref{e:decreasingsets} by
\[
\mathcal{X}_t = \{ \bm{x}\in\mathcal{X}:
S(\bm{x})\geq m_t\}.
\]
Note that in this way
\[
\mathcal{X}_t =  \{\bm{x}\in\mathcal{X}_{t-1} :
S(\bm{x})\geq m_t\},
\]
for $t=1,2,\ldots$.
The latter representation is most useful since
it shows that the ratio factor $c_t$ in \eqref{e:ratiofactor} can be
considered as a conditional expectation:
\begin{equation}\label{e:condexp}
c_t  =  \frac{|\mathcal{X}_t|}{|\mathcal{X}_{t-1}|}= \Em_{g_{t-1}} [
I_{\{ S(\bm{X}) \ge m_t \}} ],
\end{equation}
where $\bm{X}\simindist g_{t-1}=\mathcal{U}(\mathcal{X}_{t-1})$.
Note that $g_{t-1}(\bm{x})$ is also obtained as a conditional pdf
by
\begin{equation}\label{e:condpdf}
g_{t-1}(\bm{x}) = f(\bm{x}|\mathcal{X}_{t-1}) =
\begin{cases}
\frac{f(\bm{x})}{f(\mathcal{X}_{t-1})}, &
\quad \bm{x}\in\mathcal{X}_{t-1};\\
0, & \quad \bm{x}\not\in\mathcal{X}_{t-1}.
\end{cases}
\end{equation}
To draw samples from the uniform pdf $g_{t-1} =
\mathcal{U}(\mathcal{X}_{t-1})$ on a complex set given implicitly,
one applies typically MCMC methods. For further details
we refer to \cite{Rubinstein10}.
\subsection{The Basic Adaptive Splitting Algorithm}\label{ss:main}
We describe here the adaptive splitting algorithm from
\cite{BotevKroese11}.
The thresholds $(m_t)$ are not given in advance,
but determined adaptively  via a simulation process.
Hence, the number $T$ of thresholds becomes a
random variable.
In fact, the $(m_t)$-thresholds should satisfy the requirements
$c_t={|\mathcal{X}_t|/|\mathcal{X}_{t-1}|}\approx\rho_t$, where
the parameters $\rho_t\in(0,1)$ are not too small, say $\rho_t\geq 0.01$,
and set in advance. We call these the splitting control
parameters. In most applications we chose these all equal, that
is $\rho_t\equiv \rho$.
\par
Consider a sample set
$[\bm{X}]_{t-1}=\{\bm{X}_1,\ldots,\bm{X}_N\}$ of $N$
random points in $\mathcal{X}_{t-1}$. That is,
all these points are uniformly distributed on $\mathcal{X}_{t-1}$.
Let $m_{t}$ be the $(1-\rho_{t-1})$-th quantile of the ordered
statistics values of the scores
$S(\bm{X}_{1}),\ldots,S(\bm{X}_{N})$.
The elite set
$[\bm{X}]_{t-1}^{\rm(e)}\subset[\bm{X}]_{t-1}$ consists
of those points of the
sample set for which $S(\bm{X}_i)\geq m_{t}$.
Let $N_t$ be the size of the elite set.
If all scores $S(\bm{X}_i)$ would be distinct,
it follows that the number of elites
$N_{t} = \lceil N \rho_{t-1} \rceil$,
where $\lceil \cdot \rceil$ denotes rounding
to the largest integer. However, dealing with a discrete
space, typically we will find more samples with
$S(\bm{X}_i)\geq m_t$. All these are added to the elite set.
Finally we remark that from \eqref{e:condpdf} it easily follows
that the elite points are distributed uniformly on
$\mathcal{X}_t$.
\par
Having an elite set in $\mathcal{X}_t$, we do two things.
First, we screen out (delete) duplicates, so that we end up with
a set of size $N_t^{\rm(s)}$ of distinct elites.
Secondly, each screened elite is the
starting point of a Markov chain simulation (MCMC method)
on $\mathcal{X}_{t}$ using a transition probability matrix $P_{t}$ with
$g_{t}=\mathcal{U}(\mathcal{X}_{t})$ as its stationary distribution.
Because the starting point is uniformly distributed, all consecutive
points on the sample path are uniformly distributed on $\mathcal{X}_t$.
Therefore, we may use all these points in the next iteration.
\par
Suppose that each sample path has length  $b_t=\lfloor N/N_t^{\rm(s)} \rfloor$,
then we get a total of $N_t^{\rm(s)}b_t\leq N$ uniform points in
$\mathcal{X}_t$. To continue with the next iteration again with
a sample set of size $N$, we choose randomly $N-N_t^{\rm(s)}b_t$
of these sample paths and extend them by one point.
Denote the new sample set by
$[\bm{X}]_{t}$, and repeat the same procedure as above.
The algorithm iterates until we find $m_t=m$, say at iteration $T$,
at which stage we stop and deliver
\begin{equation}\label{e:basicestimator}
 \widehat{|\mathcal{X}^*|} =
|\mathcal{X}|\, \prod_{t=1}^T \hat{c}_t
\end{equation}
as an estimator of $|\mathcal{X}^*|$, where
$\hat{c}_t=N_t/N$ in iteration $t$.
\par\bigskip\noindent
In our experiments
we applied a Gibbs sampler to implement
the MCMC simulation for obtaining uniformly distributed
samples. To summarize, we give the algorithm.
\begin{alg}[Basic splitting algorithm for counting] \label{a:basic}
\mbox{}
\begin{enumerate}[1.]
\item
Set a counter $t=1$. Generate a sample set $[\bm{X}]_{0}$
of $N$ points uniformly distributed in $\mathcal{X}_0$.
Compute the threshold $m_1$, and determine the size $N_1$ of the elite set.
Set $\hat{c}_1=N_1/N$ as an estimator of $c_1=|\mathcal{X}_1|/|\mathcal{X}_0|$.
\item
Screen out the elite set to obtain $N_t^{\rm(s)}$ distinct points
uniformly distributed in $\mathcal{X}_t$.
\item
Let $b_t=\lfloor N/N_t^{\rm(s)} \rfloor$.
For all $i=1,2,\ldots,N_t^{\rm(s)}$, starting at the $i$-th screened elite point
run a Markov chain of length $b_t$ on $\mathcal{X}_t$ with
$g_{t}=\mathcal{U}(\mathcal{X}_{t})$ as its stationary distribution.
Extend $N-N_t^{\rm(s)}b_t$ randomly chosen sample paths with one point.
Denote the new sample set of size $N$ by $[\bm{X}]_{t}$.
\item
Increase the counter $t=t+1$.
Compute the threshold $m_t$, and determine the size $N_t$ of the elite set.
Set $\hat{c}_t=N_t/N$ as an estimator of $c_t=|\mathcal{X}_t|/|\mathcal{X}_{t-1}|$.
\item
If $m_t=m$ deliver the estimator \eqref{e:basicestimator};
otherwise repeat from step 2.
\end{enumerate}
\end{alg}


\section{Combining Splitting and Capture--Recapture}\label{s:CapRecap}
In this section we discuss how to combine the
well known capture-recapture (CAP-RECAP) method with the
basic splitting Algorithm \ref{a:basic}.
First we present the classical capture-cecapture algorithm in the literature.
\subsection{The Classic Capture--Recapture in the Literature}\label{ss:classic}
Originally the capture-recapture method was used to estimate the
size, say $M$, of an unknown population on the basis of
two independent samples from it.
To see how the CAP-RECAP method works, consider an urn model
with a total of $M$ identical balls. Denote by $N_1$ and $N_2$ the
sample sizes taken at the first and second draws, respectively. Assume in
addition that
\begin{itemize}
\item
The second draw takes place after all $N_1$ balls have been returned to the urn.
\item
Before returning the $N_1$ balls, each is marked, say we
painted them a different color.
\end{itemize}
Denote by $R$ the number of balls from the first draw that
reappear in the second. Then an (biased) estimate $\widetilde M$ of $M$ becomes
\[
\widetilde{M}= \frac{N_1 N_2}{R}.
\]
This is based on the observation that $N_2/M \approx R/N_1$.
Note that the name capture-recapture was borrowed from a
problem of estimating the animal population
size in a particular area on the basis of two visits.
In this case $R$ denotes the number of animals captured on the
first visit and recaptured on the second.
\par
A slightly less biased estimator of $M$ is
\begin{equation}
\widehat{M}= \frac{(N_1+1) (N_2 +1)}{(R+1)} -1. \label{e:chapmanest}
\end{equation}
See \cite{Seber70} for an analysis of the bias and for the derivation
of an approximate unbiased estimator of the variance of $\widehat{M}$:
\begin{equation}\label{e:varchapmanest}
\EE\left[\frac{(N_1+1) (N_2 +1)(N_1 -R)(N_2 -R)}{(R+1)^2(R+2)}\right]\approx
\Var(\widehat{M}).
\end{equation}
\subsection{Splitting algorithm combined with Capture--Recapture}
\label{ss:splitCapRecap}
Application of the CAP-RECAP to counting problems is trivial. We set
$|\mathcal{X}^*| =M$ and note that $N_1$ and $N_2$ correspond to the
screened-out samples at the first and second draws, which are
performed after Algorithm \ref{a:basic} reaches the desired level
$m$. Note that we need to remove duplicate samples because these
do not occur in the capture-recapture method.
\par
As an example, let us assume that we run the splitting algorithm
\ref{a:basic} till its last step $T$ with $N = 10,000$. After reaching the
desired level $m$, we draw two independent sets of magnitude $N_1 = 5000$
and $N_2 = 5010$ and assume that the number of solutions that appeared in
both draws in $10$, i.e. $R=10$. The CAP-RECAP estimator
of $|\mathcal{X}^*|$, denoted by $\widehat{|\mathcal{X}^*|}_{\rm cap}$ is
therefore
\[
\widehat{|\mathcal{X}^*|}_{\rm cap} = 2,505,000.
\]
Our numerical results in Section \ref{s:numerical} clearly indicate that the CAP-RECAP
estimator is typically more accurate than the product estimator
\eqref{e:basicestimator}, that is
\[
\Var[\widehat{|\mathcal{X}^*|}_{\rm cap}] \leq \Var[\widehat{|\mathcal{X}^*|}],
\]
provided the sample $N$ is limited, say by  $10,000$ and
$|\mathcal{X}^*|$ is large but also limited, say by  $10^6$.
\par
We make a distinction for larger solution sets:
if $10^6<|\mathcal{X}^*|\leq 10^9$, we apply an extended version of
the capture-recapture method, as we will describe in the next section.
If $|\mathcal{X}^*|$ is even larger ($|\mathcal{X}^*|>10^9$),
we can estimate it with the crude Monte Carlo.
\subsection{Extended Capture--Recapture Method}\label{ss:extended}
Recall that the regular CAP-RECAP method
\begin{enumerate}
\item
Is implemented at the last iteration $T$ of the splitting
algorithm, that is when some configurations have already reached the
desired set $\mathcal{X}^*$.
\item
It provides reliable estimators of $|\mathcal{X}^*|$ if it is
not too large, say $|\mathcal{X}^*| \leq 10^6$.
\end{enumerate}
In typical rare events counting problems, like SAT
$|\mathcal{X}^*|$ is indeed $\leq 10^6$, nevertheless we present below an
extended CAP-RECAP version, which extends the original
CAP-RECAP for $2$-$3$ orders more, that is it provides reliable
counting estimators for $10^6 < |\mathcal{X}^*| \leq 10^9$.
\par
If not stated otherwise we shall have in mind a SAT problem. The
enhanced CAP-RECAP algorithm involves additional constraints
(clauses) and can be written as follows.

\begin{alg}[Extended  CAP-RECAP] \label{a:x-cap-recap}
As soon as all $m$ clauses $C_{1}, \ldots, C_{m}$ of $\mathcal{X}_m$ have been
reached by the splitting algorithm and it occurs that the
resulting product estimator $\widehat{|\mathcal{X}_m|}$ of
$|\mathcal{X}_m|$ is larger than $ >10^6$ proceed as follows:

\begin{enumerate}[1.]
\item
Generate a sample $\bm{X}_1, \ldots, \bm{X}_{N_{\mathcal{X}_m}}$
of uniformly distributed points in the desired problem set $\mathcal{X}_m$ by adding
one by one some arbitrary auxiliary clauses
until for some $\tau$ we have that
\begin{equation}\label{e:backwardconditional}
\widehat{c_ {m+\tau}}={N_{ \mathcal{X}_{m+\tau}}\over N_{\mathcal{X}_m}} \leq
c_{m+\tau}.
\end{equation}
Here $c_ {m+\tau}$ is a relatively small number, fixed in advance,
say $10^{-2} \leq c_ {m+\tau} \leq 10^{-3}$;
furthermore, $N_{\mathcal{X}_m}$ and $N_{\mathcal{X}_{m+\tau}}$ represent
the respective number of points generated at $\mathcal{X}_m$ and accepted
at ${\mathcal{X}_{m+\tau} }$. Note that the estimate
$\widehat{c_ {m+\tau}}$ is obtained as in Step 4 of the basic
splitting algorithm \ref{a:basic}.
\item  Estimate $|\mathcal{X}^*|= |\mathcal{X}_m|$ by
\begin{equation} \label{e:backwardest}
\widehat{|\mathcal{X}_{m} |}_{\rm ecap}=
\widehat{c_{m+\tau}}^{-1} \cdot \widehat{|\mathcal{X}_{m+\tau} |}_{\rm cap}.
\end{equation}
\end{enumerate}
\end{alg}

\bigskip\noindent
We call  $\widehat{|\mathcal{X}_{m} |}_{\rm ecap}$ the
extended CAP-RECAP estimator. It is essential to bear in mind that
\begin{itemize}
\item
$\widehat{|\mathcal{X}_{m+\tau} |}_{\rm cap}$ is a CAP-RECAP estimator
rather than a splitting (product) one.
\item
$\widehat{|\mathcal{X}_{m} |}_{\rm ecap}$ does not contain the original
estimators $\widehat{c}_{1}, \ldots, \widehat{c}_{T}$ generated by the
splitting method.
\item
Since we only need here the uniformity of the samples at $\mathcal{X}_m$, we
can run the splitting method of Section \ref{ss:main} all the way with relatively small
values of sample size $N$ and splitting control parameter
$\rho$ until it reaches the vicinity of $\mathcal{X}_m$
denoted by $\mathcal{X}_{m - r}$, where $r$ is a small integer say, $r=1$
or $r=2$; and then switch to larger $N$ and $\rho$.
\item
In contrast to the splitting estimator which employs a
product of $T$ terms, formula \eqref{e:backwardest} employs only a single
$c$ factor. Recall that this additional
$\widehat{c_{m+\tau}}^{-1}$ factor allows to enlarge the CAP-RECAP estimators
of $|\mathcal{X}_m|$ for about two-three additional orders, namely from
$|\mathcal{X}_m|\approx 10^6$ to $ |\mathcal{X}_m| \approx 10^9$.
\end{itemize}
%
\section{Numerical Results} \label{s:numerical}
Below we present numerical results with the splitting algorithm  for counting.
In particular we consider the
following problems:
\begin{enumerate}
\item The 3-satisfiability problem (3-SAT)
\item Graphs with prescribed degrees
\item Contingency tables
\end{enumerate}
For the 3-SAT problem we shall also use the the CAP-RECAP method. We shall show
that  typically CAP-RECAP outperforms  the splitting algorithm.
We shall use the following notations.

\begin{notation}\label{n:notation}
For iteration $t=1,2,\ldots$
\begin{itemize}
\item
$N_t$ and $N^{\rm(s)}_t$ denote  the  actual number of elites
and the number after screening, respectively;
\item
$m^*_t$ and $m_{*t}$ denote the upper and the lower elite
levels  reached, respectively (the $m_{*t}$ levels are the same as the $m_t$ levels
in the description of the algorithm);
\item
$\rho_t$ is the splitting control parameter (we chose $\rho_t\equiv \rho$);
\item
$\hat{c}_t=N_t/N$ is the estimator of the $t$-th conditional probability;
\item
product estimator
$\widehat{|\mathcal{X}^*_t|} =|\mathcal{X}| \prod_{i=1}^t\hat{c}_i$.
\end{itemize}
\end{notation}
\subsection{The 3-Satisfiability Problem (3-SAT)}\label{ss:sat}
There are $m$ clauses of length $3$ taken from $n$ boolean (or
binary) variables $x_1,\ldots,x_n$. A literal of the $j$-th variable
is either TRUE $(x_j = 1)$ or FALSE $(x_j = 0 \Leftrightarrow
\bar{x}_j = 1$, where $\bar{x}_j = \mathrm{NOT}(x_j))$. A clause is a
disjunction of literals. We assume that all clauses consist of $3$
literals. The 3-SAT problem is defined as the problem of determining
if the variables $\bm{x} = (x_1,\ldots,x_n)$ can be assigned in a
such way as to make all clauses TRUE. More formally, let
$\mathcal{X} = \{0,1\}^n$ be the set of all configurations of the
$n$ variables, and let $C_i : \mathcal{X} \rightarrow {\{0,1\}}$, be
the $m$ clauses. Then define $\phi : \mathcal{X} \rightarrow
\{0,1\}$ by
\[
\phi(\bm{x}) = \bigwedge_{i=1}^{m} C_i(\bm{x}).
\]
The original 3-SAT problem is to find a configuration of the $x_j$
variables for which $\phi(\bm{x}) = 1$. In this work we are
interested in the total number of such configurations (or feasible
solutions). Then as discussed in Section \ref{s:split},
$\mathcal{X}^*$ denotes the set of feasible solutions. Trivially,
there are $|\mathcal{X}| = 2^n$ configurations.
\par
The $3$-SAT problems can also be converted into the family of
decision problems \eqref{e:ipset} given in Section
\ref{s:split}. Define the $m \times n$ matrix $\bm{A}$ with
entries $a_{ij} \in \{-1,0,1\}$ by
\[
a_{ij} =
\begin{cases}
-1 & \quad \mbox{if } \bar{x}_j \in C_i, \\
0  & \quad \mbox{if } x_j \not \in C_i \mbox{ and }
           \bar{x}_j \not \in C_i, \\
1 & \quad \mbox{if } x_j \in C_i.
\end{cases}
\]
Furthermore, let $\bm{b}$ be the $m$-(column) vector with entries
$b_i = 1-|\{j : a_{ij} = -1 \}|$. Then it is easy to see that for
any configuration $\bm{x} \in \{0,1\}^n$
\[
\bm{x} \in \mathcal{X}^* \Leftrightarrow \phi(\bm{x}) = 1
\Leftrightarrow \bm{Ax} \ge \bm{b}.
\]
Below we compare the efficiencies of the classic and the extended
CAP-RECAP with their splitting counterpart, bearing in mind that the
extended CAP-RECAP version is used for larger values of $|\mathcal{X}^*|$
then the classic one. As an example we consider the estimation of
$|\mathcal{X}^*|$ for the $3$-SAT problem with an instance
matrix $\bm{A}$ of
dimension $(122 \times 515)$, meaning $n=122, m=515$.
In particular Table \ref{tab1-adam} presents the the performance of the splitting
Algorithm \ref{a:basic} based on 10 independent runs using $N=25,000$ and $\rho=0.1$, while
Table \ref{tab5-adam} shoews the dynamics of a run of the
Algorithm \ref{a:basic} for the same data.

\begin{table}[H]
\caption{Performance of splitting algorithm for the 3-SAT
$(122 \times 515)$ model with  $N=25,000$ and $\rho=0.1$. }
{\small
\begin{center}
\begin{tabular}{c|c|c|c}
Run & nr.\ of its. & $\widehat{|\mathcal{X}^*|}$   & CPU  \\
\hline \hline
1	&	33	&	1.41E+06	&	 212.32	\\	\hline
2	&	33	&	1.10E+06	&	 213.21	\\	\hline
3	&	33	&	1.68E+06	&	 214.05	\\	\hline
4	&	33	&	1.21E+06	&	 215.5	\\	\hline
5	&	33	&	1.21E+06	&	 214.15	\\	\hline
6	&	33	&	1.47E+06	&	 216.05	\\	\hline
7	&	33	&	1.50E+06	&	 252.25	\\	\hline
8	&	33	&	1.73E+06	&	 243.26	\\	\hline
9	&	33	&	1.21E+06	&	 238.63	\\	\hline
10	&	33	&	1.88E+06	&	 224.36	\\	\hline
\hline								
Average	&	33	&	1.44E+06	&	 224.38	\\ \hline
\end{tabular}
\end{center}
}
\label{tab1-adam}
\end{table}

The relative error, denoted by RE is $1.815E-01$.
Notice that the relative error of a random variable $Z$ is calculated by the standard formula, namely

\[ RE = {S /\widehat {\ell}},\]
 where
\[ \widehat \ell ={1\over N}\sum_{i=1}^N Z_i, \quad
S^2 = {1\over N-1}\sum_{i=1}^N (Z_i - \widehat \ell)^2.
\]

\begin{table}[H]
\caption{Dynamics of a run of the splitting
algorithm  for the 3-SAT
$(122 \times 515)$ model using $N=25,000$ and $\rho=0.1$.}
{\small
\begin{center}
\begin{tabular}{c|c|c|c|c|c|c}
$t$ & $\widehat{|\mathcal{X}^*_t|}$  & $N_t$ & $N_t^{\rm(s)}$ &
$m_{t}^{*}$ & $m_{*t}$ & $\hat{c}_{t}$ \\
\hline
1	&	6.53E+35	&	3069	&	 3069	&	480	&	460	&	 1.23E-01	\\
2	&	8.78E+34	&	3364	&	 3364	&	483	&	467	&	 1.35E-01	\\
3	&	1.15E+34	&	3270	&	 3270	&	484	&	472	&	 1.31E-01	\\
4	&	1.50E+33	&	3269	&	 3269	&	489	&	476	&	 1.31E-01	\\
5	&	2.49E+32	&	4151	&	 4151	&	490	&	479	&	 1.66E-01	\\
6	&	3.37E+31	&	3379	&	 3379	&	492	&	482	&	 1.35E-01	\\
7	&	3.41E+30	&	2527	&	 2527	&	494	&	485	&	 1.01E-01	\\
8	&	6.19E+29	&	4538	&	 4538	&	495	&	487	&	 1.82E-01	\\
9	&	9.85E+28	&	3981	&	 3981	&	497	&	489	&	 1.59E-01	\\
10	&	1.31E+28	&	3316	&	 3316	&	498	&	491	&	 1.33E-01	\\
11	&	1.46E+27	&	2797	&	 2797	&	501	&	493	&	 1.12E-01	\\
12	&	4.61E+26	&	7884	&	 7884	&	501	&	494	&	 3.15E-01	\\
13	&	1.36E+26	&	7380	&	 7380	&	501	&	495	&	 2.95E-01	\\
14	&	3.89E+25	&	7150	&	 7150	&	502	&	496	&	 2.86E-01	\\
15	&	1.06E+25	&	6782	&	 6782	&	505	&	497	&	 2.71E-01	\\
16	&	2.69E+24	&	6364	&	 6364	&	503	&	498	&	 2.55E-01	\\
17	&	6.42E+23	&	5969	&	 5969	&	504	&	499	&	 2.39E-01	\\
18	&	1.42E+23	&	5525	&	 5525	&	506	&	500	&	 2.21E-01	\\
19	&	3.03E+22	&	5333	&	 5333	&	505	&	501	&	 2.13E-01	\\
20	&	5.87E+21	&	4850	&	 4850	&	506	&	502	&	 1.94E-01	\\
21	&	1.06E+21	&	4496	&	 4496	&	507	&	503	&	 1.80E-01	\\
22	&	1.71E+20	&	4061	&	 4061	&	507	&	504	&	 1.62E-01	\\
23	&	2.50E+19	&	3647	&	 3647	&	509	&	505	&	 1.46E-01	\\
24	&	3.26E+18	&	3260	&	 3260	&	510	&	506	&	 1.30E-01	\\
25	&	3.62E+17	&	2778	&	 2778	&	510	&	507	&	 1.11E-01	\\
26	&	3.68E+16	&	2539	&	 2539	&	510	&	508	&	 1.02E-01	\\
27	&	3.05E+15	&	2070	&	 2070	&	511	&	509	&	 8.28E-02	\\
28	&	2.17E+14	&	1782	&	 1782	&	512	&	510	&	 7.13E-02	\\
29	&	1.21E+13	&	1398	&	 1398	&	513	&	511	&	 5.59E-02	\\
30	&	5.00E+11	&	1030	&	 1030	&	513	&	512	&	 4.12E-02	\\
31	&	1.49E+10	&	743	&	743	 &	514	&	513	&	2.97E-02	\\
32	&	2.39E+08	&	402	&	402	 &	515	&	514	&	1.61E-02	\\
33	&	1.43E+06	&	150	&	150	 &	515	&	515	&	6.00E-03	
\end{tabular}
\end{center}
}
\label{tab5-adam}
\end{table}

We increased the sample size
at the last two iterations from $N=25,000$ to $N=100,000$
to get a more accurate estimator.
\par\bigskip\noindent
As can be seen from Table \ref{tab1-adam}, the estimator
$\widehat{|\mathcal{X}^*|} >10^6$, hence for this instance the extended CAP-RECAP
Algorithm \ref{a:x-cap-recap} can also be used. We shall show that
the relative error (RE) of the extended CAP-RECAP estimator
$\widehat{|\mathcal{X}_{m} |}_{\rm ecap}$ is less than that of
$\widehat{|\mathcal{X}^*|}$. Before doing so we
need to find the extended 3-SAT instance matrix $(122\times 515)+\tau$,
where $\tau$ is the number of auxiliary clauses.
Applying the extended CAP-RECAP Algorithm \ref{a:x-cap-recap} we
found that $\tau =5$ and thus the extended instance matrix is
$(122\times 520)$. Recall that the cardinality $|\mathcal{X}_{m+\tau}|$ of
the extended  $(122\times 520)$ model should be manageable by the
regular CAP-RECAP, that is we assumed that $|\mathcal{X}_{m+\tau}|< 10^6$.
Indeed, Table \ref{tab2-adam} presents the performance of the
regular CAP-RECAP for that extended $(122\times 520)$ model. Here we
used again $\rho=0.1$. As for the sample size, we took $N=1,000$
until iteration $28$ and then switched to $N=100,000$. The final
CAP-RECAP estimator is obtained by taking two equal samples, each of
size  $N=70,000$ at the final subset
$\mathcal{X}_{m+\tau} = \mathcal{X}_{520}$. (The sample sizes that
were used in the estimation are smaller due to the screening step.)

\begin{table}[H]
\caption{Performance of the regular CAP-RECAP for the extended
$(122\times 520)$ model.}
{\small
\begin{center}
\begin{tabular}{c|c|c|c}
Run & nr.\ of its.  & $\widehat{|\mathcal{X}^*|}_{\rm cap}$
&  CPU  \\
\hline \hline
1	&	34	&	5.53E+04	&	 159.05	\\	\hline	
2	&	35	&	5.49E+04	&	 174.46	\\	\hline	
3	&	35	&	5.51E+04	&	 178.08	\\	\hline	
4	&	34	&	5.51E+04	&	 166.36	\\	\hline	
5	&	34	&	5.52E+04	&	 159.36	\\	\hline	
6	&	33	&	5.52E+04	&	 152.38	\\	\hline	
7	&	33	&	5.54E+04	&	 137.96	\\	\hline	
8	&	34	&	5.50E+04	&	 157.37	\\	\hline	
9	&	35	&	5.51E+04	&	 179.08	\\	\hline	
10	&	34	&	5.51E+04	&	 163.7	\\	\hline	\hline					
Average	&	34.1	&	5.51E+04	 &	162.78	\\	\hline	

\end{tabular}
\end{center}
}
\label{tab2-adam}
\end{table}

The relative error of  $\widehat{|\mathcal{X}^*|}_{\rm cap}$ over
$10$ runs is $2.600E-03$.
\par\bigskip\noindent
Next we compare the efficiency of the regular CAP-RECAP  (as per
Table \ref{tab2-adam}) with  that of the splitting algorithm for the
extended  $(122\times 520)$ model. Table \ref{tab3-adam} presents
the performance of splitting for $\rho=0.1$ and $N=100,000$. It
readily follows that the relative error of the regular CAP-RECAP is
about $30$ times less than that of splitting. Notice in addition
that the CPU time of CAP-RECAP is about $6$ times less than that of
splitting. This is so since the total sample size of the former is about
$6$ time less than of the latter. Thus the overall speed up obtained
by CAP-RECAP is about $5,000$ times.

\begin{table}[H]
 \caption{ Performance of splitting algorithm for the 3-SAT
 $(122 \times 520)$ model.}
{\small
\begin{center}
\begin{tabular}{c|c|c|c}
Run & nr.\ of its.  & $\widehat{|\mathcal{X}^*|}$
   & CPU  \\
\hline \hline
1	&	34	&	6.03E+04	&	 900.28	\\	\hline	
2	&	34	&	7.48E+04	&	 904.23	\\	\hline	
3	&	34	&	4.50E+04	&	 913.31	\\	\hline	
4	&	34	&	5.99E+04	&	 912.27	\\	\hline	
5	&	34	&	6.03E+04	&	 910.44	\\	\hline	
6	&	33	&	4.94E+04	&	 898.91	\\	\hline	
7	&	34	&	5.22E+04	&	 931.88	\\	\hline	
8	&	34	&	5.74E+04	&	 916.8	\\	\hline	
9	&	34	&	5.85E+04	&	 919.63	\\	\hline	
10	&	34	&	5.72E+04	&	 927.7	\\	\hline	\hline
Average	&	33.9	&	5.75E+04	 &	913.54	\\	\hline	
\end{tabular}
\end{center}
}
\label{tab3-adam}
\end{table}

The relative error of  $\widehat{|\mathcal{X}^*|}$ over $10$ runs is $1.315E-01$.
\par\bigskip\noindent
With these results at hand we can proceed with the extended
CAP-RECAP and compare its efficiency with splitting (see Table
\ref{tab1-adam}) for the instance matrix $(122 \times 515)$. Table
\ref{tab4-adam} presents the performance of the extended
CAP-RECAP estimator $\widehat{|\mathcal{X}^*|}_{\rm ecap}$ for the
$(122 \times 515)$ model along with the performance of the regular
CAP-RECAP estimator $\widehat{|\mathcal{X}^*|}_{\rm cap}$ for the $(122
\times 520)$ model (see also the results of Table \ref{tab2-adam}
for $\widehat{|\mathcal{X}^*|}_{\rm cap}$). We set again $\rho=0.1$.
Regarding the sample size we took $N=1,000$ for the first $31$
iterations and then switched to $N=100,000$ until reaching the level
$m=515$. Recall that the level $m+\tau =520$ and the corresponding
CAP-RECAP estimator $\widehat{|\mathcal{X}^*|}_{\rm cap}$ was obtained
from the set $\mathcal{X}_{m} =\mathcal{X}_{515}$ by
 adding $\tau =5$ more auxiliary clauses. Note that in this
case we used  for $\widehat{|\mathcal{X}^*|}_{\rm cap}$ two equal samples
each of length $N = 100,000$.
\par
Comparing the results of Table \ref{tab1-adam} with that of Table
\ref{tab4-adam} it is readily seen that the extended CAP-RECAP
estimator $\widehat{|\mathcal{X}^*|}_{\rm ecap}$ outperforms the splitting
one $\widehat{|\mathcal{X}^*|}$ in both RE and CPU time. In particular,
we have that both RE and CPU times of the former are about $1.6$ times
less than of the latter. This means that the overall speed up
obtained by $\widehat{|\mathcal{X}^*|}_{\rm ecap}$ versus
$\widehat{|\mathcal{X}^*|}$ is about $1,6^2 \cdot 1.6 \approx 4$ times. Note
finally that the total number of samples used in the extended
CAP-RECAP estimator $\widehat{|\mathcal{X}^*|}_{\rm ecap}$ is about
$N=500,000$, while in its counterpart - the splitting estimator
$\widehat{|\mathcal{X}^*|}$ is about $N= 50,000 * 36 = 1,800,000$.

\begin{table}[H]
\caption{Performance of the extended CAP-RECAP estimator
$\widehat{|\mathcal{X}^*|}_{\rm ecap}$ for the $(122 \times 515)$ model
along with the regular CAP-RECAP one $\widehat{|\mathcal{X}^*|}_{\rm cap}$
for the $(122 \times 520)$ model.}
{\small
\begin{center}
\begin{tabular}{c|c|c|c|c|c}
Run & nr.\ its. & $\widehat{c_{m+\tau}}$
& $\widehat{|\mathcal{X}^*|}_{\rm cap}$ &  $\widehat{|\mathcal{X}^*|}_{\rm ecap}$  &
 {CPU}   \\
\hline \hline
1	&	33	&	3.13E-02	&	 5.41E+04	&	1.73E+06	&	 138.99	\\	\hline		
2	&	34	&	3.47E-02	&	 5.51E+04	&	1.59E+06	&	 154.64	\\	\hline		
3	&	34	&	3.55E-02	&	 5.52E+04	&	1.55E+06	&	 161.78	\\	\hline		
4	&	33	&	4.51E-02	&	 5.40E+04	&	1.20E+06	&	 163.53	\\	\hline		
5	&	34	&	3.04E-02	&	 5.13E+04	&	1.69E+06	&	 143.84	\\	\hline		
6	&	34	&	2.99E-02	&	 5.41E+04	&	1.81E+06	&	 151.1	\\	\hline		
7	&	34	&	4.27E-02	&	 5.51E+04	&	1.29E+06	&	 174.08	\\	\hline		
8	&	34	&	3.87E-02	&	 5.42E+04	&	1.40E+06	&	 143.27	\\	\hline		
9	&	33	&	3.27E-02	&	 5.42E+04	&	1.66E+06	&	 171.07	\\	\hline		
10	&	34	&	4.22E-02	&	 5.51E+04	&	1.30E+06	&	 154.71	\\	\hline	\hline	
Average	&	33.7	&	3.63E-02	 &	5.42E+04	&	1.52E+06	&	 155.70\\	\hline			
\end{tabular}
\end{center}
}
\label{tab4-adam}
\end{table}

The relative error of  $\widehat{|\mathcal{X}^*|}_{\rm cap}$ over $10$ runs is $2.010E-02$.

The relative error of  $\widehat{|\mathcal{X}^*|}_{\rm ecap}$ over $10$ runs is $1.315E-01$.
\subsection{Random graphs with prescribed degrees}\label{ss:graphs}
Random graphs with given vertex degrees have attained attention as a
model for real-world complex networks, including World Wide Web,
social networks and biological networks. The problem is basically
finding a graph $G=(V,E)$ with $n$ vertices, given the degree
sequence $\bm{d} = (d_1,\ldots,d_n)$ formed of nonnegative integers.
Following \cite{Blitzstein06}, a finite sequence $(d_1,\ldots,d_n)$
of nonnegative integers is called graphical if there is a labeled
simple graph with vertex set $\{ 1,\ldots,n\}$ in which vertex $i$
has degree $d_i$. Such a graph is called a realization of the degree
sequence $(d_1,\ldots,d_n)$. We are interested in
the total number of realizations for a given degree sequence, hence
$\mathcal{X}^*$ denotes the set of all graphs $G=(V,E)$ with the
degree sequence $(d_1,\ldots,d_n)$.
\par
Similar to \eqref{e:ipset} for SAT we convert the problem into
a  counting  problem. To proceed
consider the complete graph $K_n$ of $n$ vertices,
in which each vertex is connected with all other
vertices. Thus the total number of edges in $K_n$
is $m=n(n-1)/2$, labeled $e_1,\ldots,e_m$.
The random graph problem with prescribed degrees is translated to the problem
of choosing those edges of $K_n$ such that the resulting graph $G$
matches the given sequence $\bm{d}$. Set $x_i=1$ when $e_i$ is chosen,
and $x_i=0$ otherwise, $i=1,\ldots,m$.
In order that such an assignment $\bm{x}\in\{0,1\}^m$
matches the given degree sequence $(d_1,\ldots,d_n)$,
it holds necessarily that $\sum_{j=1}^mx_j=\tfrac12\sum_{i=1}^nd_i$,
since this is the total number of edges.
In other words,
the configuration space is
\[
\mathcal{X}=\left\{ \bm{x}\in\{0,1\}^m:
\sum_{j=1}^mx_j=\tfrac12\sum_{i=1}^nd_i\right\}.
\]
Let $\bm{A}$ be the incidence matrix of $K_n$ with entries
\[
a_{ij} =
\begin{cases}
0 & \quad \mbox{if } v_i \not \in e_j \\
1 & \quad \mbox{if } v_i \in e_j.
\end{cases}
\]
It is easy to see that whenever a configuration $\bm{x}\in\{0,1\}^m$
satisfies $\bm{Ax} = \bm{d}$, the associated graph has
degree sequence $(d_1,\ldots,d_n)$. We conclude that the
problem set is represented by
\[
\mathcal{X}^* = \{\bm{x}\in\mathcal{X}: \bm{Ax} = \bm{d}\}.
\]
We first present a  small example as illustration.
Let  $\bm{d}=(2,2,2,1,3)$ with $n=5$, and $m=10$.
After ordering the edges of $K_5$ lexicographically,
the corresponding incidence matrix is given as
\[
\bm{A} =
\begin{pmatrix}
1 & 1 & 1 & 1 & 0 & 0 & 0 & 0 & 0 & 0 \\
1 & 0 & 0 & 0 & 1 & 1 & 1 & 0 & 0 & 0\\
0 & 1 & 0 & 0 & 1 & 0 & 0 & 1 & 1 & 0\\
0 & 0 & 1 & 0 & 0 & 1 & 0 & 1 & 0 & 1 \\
0 & 0 & 0 & 1 & 0 & 0 & 1 & 0 & 1 & 1
\end{pmatrix}
\]
 It is readily seen that the following
$\bm{x}=(0,0,1,1,1,0,1,0,1,0)'$, and
$\bm{x}=(1,0,0,1,1,0,0,0,1,1)'$  present two  solutions of this example.
\par
For the random graph problem we define the score function
$S : \mathcal{X}\rightarrow \mathbb{Z}_-$ by
\[
S(\bm{x}) = -\sum_{i=1}^{n} |\mathrm{deg}(v_i) - d_i|,
\]
where $\mathrm{deg}(v_i)$ is the degree of vertex $i$ under
the configuration $\bm{x}$.
Each configuration that satisfies the degree sequence will have
a performance function equal to $0$.
\par
The implementation of the Gibbs sampler for this problem is slightly
different than for the $3$-SAT problem, since we keep the
number of edges in each realization fixed to $\sum d_i /2$.
Our first algorithm takes care of this requirement
and generates a random $\bm{x}\in\mathcal{X}$.

\begin{alg}
Let $(d_1,\ldots,d_n)$ be the prescribed degrees sequence.
\begin{itemize}
\item
Generate a random permutation of $1,\ldots,m$.
\item
Choose the first $\sum d_i/2$ places in this permutation and
deliver a vector $\bm{x}$ having one's in those places.
\end{itemize}
\end{alg}

The adaptive thresholds in the basic splitting algorithm
are negative, increasing to 0:
\[
m_1\leq m_2\leq\cdots\leq m_{T-1}\leq m_T=0.
\]
 The resulting Gibbs sampler (in Step 3 of the basic splitting algorithm starting
with a configuration $\bm{x}\in\mathcal{X}$ for which
$S(\bm{x})\geq m_t$)  can be written as follows.

\begin{alg}[Gibbs Algorithm for random graph sampling]
For each edge $x_i=1$, while keeping all other edges fixed, do:
\begin{enumerate}[1.]
\item
Remove $x_i$ from $\bm{x}$, i.e. make $x_i = 0$.
\item
Check all possible placements for the edge resulting a
new vector $\bar{\bm{x}}$ conditioning on the performance function
$S(\bar{\bm{x}}) \geq m_t$
\item
With uniform probability choose one of the valid realizations.
\end{enumerate}
\end{alg}

We will apply the splitting algorithm to two problems taken
from \cite{Blitzstein06}.

\subsubsection{A small problem}
For this small problem we have the degree sequence
\[
\bm{d} = (5, 6, \underbrace{1, \ldots ,1}_{\mbox{11 ones}}).
\]
The solution
can be obtained analytically and already given in \cite{Blitzstein06}:
\begin{quote}
``To count the number of labeled graphs with this degree sequence,
note that there are ${11 \choose 5}=462$  such graphs with vertex 1
not joined to vertex 2 by an edge (these graphs look like two separate stars),
and there are ${11 \choose 4}{7 \choose 5} = 6930$ such graphs with an
edge between vertices 1 and 2 (these look like two joined stars with
an isolated edge left over). Thus, the total number of realizations
of $\bm{d}$ is $7392$.''
\end{quote}
As we can see from Table \ref{tab1}, the algorithm easily handles the problem.
Table \ref{tab2} presents the typical dynamics.

\begin{table}[H]
\caption{Performance of the splitting algorithm for a small problem
using $N=50,000$ and $\rho=0.5$.}
{\small
\begin{center}
\begin{tabular}{c|c|c|c}
Run & nr.\ of its.  & $\widehat{|\mathcal{X}^*|}$ &
 CPU  \\
\hline \hline
1	&	10	&	7146.2	&	15.723	 \\	\hline	
2	&	10	&	7169.2	&	15.251	 \\	\hline	
3	&	10	&	7468.7	&	15.664	 \\	\hline	
4	&	10	&	7145.9	&	15.453	 \\	\hline	
5	&	10	&	7583	&	15.555	 \\	\hline	
6	&	10	&	7206.4	&	15.454	 \\	\hline	
7	&	10	&	7079.3	&	15.495	 \\	\hline	
8	&	10	&	7545.1	&	15.347	 \\	\hline	
9	&	10	&	7597.2	&	15.836	 \\	\hline	
10	&	10	&	7181.2	&	15.612	 \\	\hline	\hline
Average	&	10	&	7312.2	&	 15.539	\\	\hline	
\end{tabular}
\end{center}
}
\label{tab1}
\end{table}

The relative error of  $\widehat{|\mathcal{X}^*|}$ over $10$ runs is $2.710E-02$.

\begin{table}[H]
\caption{Typical dynamics of the splitting algorithm for
a small problem using $N=50,000$ and $\rho=0.5$
(recall Notation \ref{n:notation} at the beginning of Section \ref{s:numerical}).}
{\small
\begin{center}
\begin{tabular}{c|c|c|c|c|c|c}
$t$ & $\widehat{|\mathcal{X}_t^*|}$  & $N_t$ & $N_t^{\rm(s)}$ &
$m_{t}^{*}$ & $m_{*t}$ & $\hat{c}_{t}$ \\
\hline
1   &   4.55E+12    &   29227   &   29227   &   -4  &   -30 &   0.5845  \\
2   &   2.56E+12    &   28144   &   28144   &   -4  &   -18 &   0.5629  \\
3   &   1.09E+12    &   21227   &   21227   &   -6  &   -16 &   0.4245  \\
4   &   3.38E+11    &   15565   &   15565   &   -4  &   -14 &   0.3113  \\
5   &   7.51E+10    &   11104   &   11104   &   -4  &   -12 &   0.2221  \\
6   &   1.11E+10    &   7408    &   7408    &   -2  &   -10 &   0.1482  \\
7   &   1.03E+09    &   4628    &   4628    &   -2  &   -8  &   0.0926  \\
8   &   5.37E+07    &   2608    &   2608    &   -2  &   -6  &   0.0522  \\
9   &   1.26E+06    &   1175    &   1175    &   0   &   -4  &   0.0235  \\
10  &   7223.9      &   286     &   280     &   0   &   -2  &   0.0057
\end{tabular}
\end{center}
}
\label{tab2}
\end{table}

\subsubsection{A large problem}
A much harder instance (see \cite{Blitzstein06}) is defined by
\[
\bm{d} = (7, 8, 5, 1, 1, 2, 8, 10, 4, 2, 4, 5, 3, 6, 7, 3, 2, 7, 6,
1, 2, 9, 6, 1, 3, 4, 6, 3, 3, 3, 2, 4, 4).
\]
In \cite{Blitzstein06} the number of such graphs is estimated
to be about $1.533 \times 10^{57}$
Table \ref{tab3} presents 10 runs using the splitting algorithm.

\begin{table}[H]
\caption{Performance of the splitting algorithm for
a large problem using $N=100,000$ and $\rho=0.5$.}
{\small
\begin{center}
\begin{tabular}{c|c|c|c}
Run & nr.\ its.  & $\widehat{|\mathcal{X}^*|}$
   & CPU  \\
\hline \hline
1	&	39	&	1.66E+57	&	 4295	\\	\hline	
2	&	39	&	1.58E+57	&	 4223	\\	\hline	
3	&	39	&	1.58E+57	&	 4116	\\	\hline	
4	&	39	&	1.53E+57	&	 4281	\\	\hline	
5	&	39	&	1.76E+57	&	 4301	\\	\hline	
6	&	39	&	1.75E+57	&	 4094	\\	\hline	
7	&	39	&	1.46E+57	&	 4512	\\	\hline	
8	&	39	&	1.71E+57	&	 4287	\\	\hline	
9	&	39	&	1.39E+57	&	 4158	\\	\hline	
10	&	39	&	1.38E+57	&	 4264	\\	\hline	\hline
Average	&	39	&	1.58E+57	&	 4253	\\	\hline	

\end{tabular}
\end{center}
}
\label{tab3}
\end{table}

The relative error of  $\widehat{|\mathcal{X}^*|}$ over $10$ runs is $8.430E-02$.
\subsection{Binary Contingency Tables}
Given are two vectors of positive integers
$\bm{r}=(r_1,\ldots,r_m)$ and $\bm{c}=(c_1,\ldots,c_n)$
such that $r_i\leq n$ for all $i$, $c_j\leq n$ for all $j$, and
$\sum_{i=1}^mr_i=\sum_{j=1}^nc_j$.
A \textit{binary contingency table} with row sums $\bm{r}$ and
column sums $\bm{c}$ is a $m\times n$ matrix $\bm{X}$
of zero-one entries $x_{ij}$ satisfying
$\sum_{j=1}^nx_{ij}=r_i$ for every row $i$ and
$\sum_{i=1}^mx_{ij}=c_j$ for every column $j$.
The problem is to count all contingency tables.
\par
The extension of the proposed Gibbs sampler for counting the
contingency tables is straightforward.
We define the configuration space
$\mathcal{X}= \mathcal{X}^{\rm(r)}\cup \mathcal{X}^{\rm(c)}$ as the
space where all column or row sums are satisfied:
\begin{align*}
& \mathcal{X}^{\rm(c)}=\left\{\bm{X}\in\{0,1\}^{m+n}: \sum_{i=1}^mx_{ij}=c_j\;\forall j\right\},\\
& \mathcal{X}^{\rm(r)}=\left\{\bm{X}\in\{0,1\}^{m+n}: \sum_{j=1}^nx_{ij}=r_i\;\forall i\right\}.
\end{align*}
Clearly we can sample uniformly at random from $\mathcal{X}$ without any problem.
The score function $S:\mathcal{X}\to\Zset_-$ is defined by
\[
S(\bm{X}) =
\begin{cases}
-\sum_{i=1}^m|\sum_{j=1}^nx_{ij} - r_i|,&\quad \text{for}\;\;\bm{X}\in \mathcal{X}^{\rm(c)},\\
 -\sum_{j=1}^n|\sum_{i=1}^mx_{ij} - c_j|,&\quad \text{for}\;\;\bm{X}\in \mathcal{X}^{\rm(r)},
\end{cases}
\]
that is, the difference of the row sums $\sum_{j=1}^nx_{ij}$ with the target $r_i$
if the column sums are right, and vice versa.
\par
The Gibbs sampler is very similar to the one in the previous
section concerning random graphs with prescribed degrees.

\begin{alg}[Gibbs algorithm for random contingency tables sampling]
Given a matrix realization $\bm{X}\in\mathcal{X}^{\rm(c)}$ with score $S(\bm{X})\geq m_t$.
For each column $j$ and for each 1-entry in this column ($x_{ij}=1$) do:
\begin{enumerate}[1.]
\item
Remove this $1$, i.e. set $x'_{ij}=0$.
\item
Check all possible placements for this $1$  in the given column
$j$ conditioning on the performance function $S(\bm{X}') \geq m_t$
($\bm{X}'$ is the matrix resulting by setting $x'_{ij}=0$,
$x'_{i'j}=1$ for some $x_{i'j}=0$, and all other entries remain unchanged).
\item
Suppose that the set of valid realization is
$\mathcal{A} = \{\bm{X}' |S(\bm{X}') \geq m_t\}$. (Please note that this set
also contains the original realization $\bm{X}$). Than with probability
$\frac{1}{|\mathcal{A}|}$ pick any realization at random and continue with
step 1.
\end{enumerate}
\end{alg}
Note that in this way we keep the column sums correct.
Similarly, when we started with a matrix configuration with all row sums correct,
we execute these steps for each row and swap 1 and 0 per row.
\subsubsection{Model 1}
The date are $m=12, n = 12$ with row and column sums
\[
\bm{r}=(2,2,2,2,2,2,2,2,2,2,2,2),\;
\bm{c}=(2,2,2,2,2,2,2,2,2,2,2,2).
\]
The true count value is known to be $21,959,547,410,077,200$.
Table \ref{tab5} presents $10$ runs using the splitting algorithm.
Table \ref{tab6} presents a typical dynamics.

\begin{table}[H]
\caption{Performance of the splitting algorithm for Model 1
using $N=50,000$ and $\rho=0.5$. }
{\small
\begin{center}
\begin{tabular}{c|c|c|c}
Run & nr.its.  & $\widehat{|\mathcal{X}^*|}$
  & CPU  \\
\hline \hline
1	&	7	&	2.15E+16	&	 4.54	\\	\hline	
2	&	7	&	2.32E+16	&	 4.55	\\	\hline	
3	&	7	&	2.23E+16	&	 4.54	\\	\hline	
4	&	7	&	2.11E+16	&	 4.58	\\	\hline	
5	&	7	&	2.05E+16	&	 4.57	\\	\hline	
6	&	7	&	2.23E+16	&	 4.54	\\	\hline	
7	&	7	&	2.02E+16	&	 4.55	\\	\hline	
8	&	7	&	2.38E+16	&	 4.58	\\	\hline	
9	&	7	&	2.06E+16	&	 4.57	\\	\hline	
10	&	7	&	2.14E+16	&	 4.55	\\	\hline	\hline
Average	&	7	&	2.17E+16	&	 4.56	\\	\hline	

\end{tabular}
\end{center}
}
\label{tab5}
\end{table}

The relative error of  $\widehat{|\mathcal{X}^*|}$ over $10$ runs is $5.210E-02$.

\begin{table}[H]
\caption{Typical dynamics of the splitting algorithm for Model 1 using
$N=50,000$ and $\rho=0.5$.}
{\small
\begin{center}
\begin{tabular}{c|c|c|c|c|c|c}
$t$ & $\widehat{|\mathcal{X}_t^*|}$  & $N_t$ & $N_t^{(s)}$ &
$m_{t}^{*}$ & $m_{*t}$ & $\hat{c}_{t}$ \\
\hline
1   &   4.56E+21    &   13361   &   13361   &   -2  &      -24 &   0.6681  \\
2   &   2.68E+21    &   11747   &   11747   &   -2  &      -12 &   0.5874  \\
3   &   1.10E+21    &   8234    &   8234    &   -2  &      -10 &   0.4117  \\
4   &   2.76E+20    &   5003    &   5003    &   -2  &      -8  &   0.2502  \\
5   &   3.45E+19    &   2497    &   2497    &   0   &      -6  &   0.1249  \\
6   &   1.92E+18    &   1112    &   1112    &   0   &      -4  &   0.0556  \\
7   &   2.08E+16    &   217 &   217 &   0   &           -2  &   0.0109
\end{tabular}
\end{center}
}
\label{tab6}
\end{table}
\subsubsection{Model  2}
Darwin's Finch Data from Yuguo Chen, Persi Diaconis, Susan P. Holmes,
and Jun S. Liu:
$m=12, n = 17$ with row and columns sums
\[
\bm{r}=(14,13,14,10,12,2,10,1,10,11,6,2),\;
\bm{c}=(3,3,10,9,9,7,8,9,7,8,2,9,3,6,8,2,2).
\]
The true count value is known to be  $67,149,106,137,567,600$.
Table \ref{tab7} presents $10$ runs using the splitting algorithm.

\begin{table}[H]
\caption{Performance of the splitting algorithm for Model 2
using $N=200,000$ and $\rho=0.5$. }
{\small
\begin{center}
\begin{tabular}{c|c|c|c}
Run  & nr. its.  & $\widehat{|\mathcal{X}^*|}$
 & CPU  \\
\hline \hline
1	&	24	&	6.16E+16	&	 246.83	\\	\hline	
2	&	24	&	6.50E+16	&	 244.42	\\	\hline	
3	&	24	&	7.07E+16	&	 252.71	\\	\hline	
4	&	24	&	7.91E+16	&	 247.36	\\	\hline	
5	&	24	&	6.61E+16	&	 260.99	\\	\hline	
6	&	24	&	6.77E+16	&	 264.07	\\	\hline	
7	&	24	&	6.59E+16	&	 269.86	\\	\hline	
8	&	24	&	6.51E+16	&	 273.51	\\	\hline	
9	&	24	&	7.10E+16	&	 272.49	\\	\hline	
10	&	24	&	5.91E+16	&	 267.23	\\	\hline	\hline
Average	&	24	&	6.71E+16	&	 259.95	\\	\hline	
\end{tabular}
\end{center}
}
\label{tab7}
\end{table}

The relative error of  $\widehat{|\mathcal{X}^*|}$ over $10$ runs is $7.850E-02$.

\section{Concluding Remarks}\label{s:conc}
In this paper we applied the splitting method to several well-known counting
problems, like 3-SAT, random graphs with
prescribed degrees and binary contingency tables.
While implementing the splitting algorithm, we discussed several MCMC algorithms and
in particular the Gibbs
sampler. We show how to incorporate the classic capture-recapture
method in the splitting algorithm in order to obtain a low
variance estimator for the  desired counting quantity. Furthermore, we presented an
extended version
of the capture-recapture algorithm, which is suitable for problems with a
larger number of feasible solutions.
We finally presented numerical results
with the splitting and capture-recapture estimators, and
showed the superiority of the latter.

\large
\renewcommand{\baselinestretch}{1.0}

\end{document}